\documentclass[a4paper,11pt]{article}
\usepackage{pos}
\usepackage{def}
\usepackage{slashed}
\usepackage[export]{adjustbox}

\title{A Brief Look at the Chirality-Flow Formalism for Standard Model Amplitudes}

\author[a]{Joakim Alnefjord}
\author*[a]{Andrew Lifson}
\author[a]{Christian Reuschle}
\author[a]{Malin Sjodahl}

\affiliation[a]{Department of Astronomy and Theoretical Physics, Lund
  University,\\ 
  S{\"o}lvegatan 14A, 223\,62 Lund, Sweden 
  }

\emailAdd{joakim.alnefjord@gmail.com}
\emailAdd{andrew.lifson@thep.lu.se}
\emailAdd{christian.reuschle@thep.lu.se}
\emailAdd{malin.sjodahl@thep.lu.se}

\abstract{
Inspired by the flow description of su(N) colour calculations,
we recently showed how to simplify the spinor-helicity formalism 
(at the algebra level two copies of complexified su(2))
by treating each Weyl spinor as part of a flow line with definite chirality and momentum.
This formalism, dubbed the chirality-flow formalism,
eliminates all non-trivial algebra from tree-level spinor-helicity calculations, 
thus allowing the shortest possible route from Feynman diagrams to complex numbers (spinor inner products). 
In this presentation, we briefly introduce the main features of this method and show some examples.

}

\FullConference{
 
 The Ninth Annual Conference on Large Hadron Collider Physics - LHCP2021\\
 7-12 June 2021\\
 Online
}

\begin{document}
\maketitle

\section{Introduction and the  massless spinor-helicity formalism}
The spinor-helicity formalism is often the most convenient framework in which to perform scattering amplitude calculations 
\cite{DeCausmaecker:1981jtq,Berends:1981rb,Berends:1981uq,DeCausmaecker:1981wzb,Berends:1983ez,Kleiss:1984dp,Berends:1984gf,Gunion:1985bp,Gunion:1985vca,Kleiss:1985yh,Hagiwara:1985yu,Kleiss:1986ct,Kleiss:1986qc,Xu:1986xb,Gastmans:1987qz,Schwinn:2005pi,Farrar:1983wk,Berends:1987cv,Berends:1987me,Berends:1988yn,Berends:1988zn,Berends:1989hf,Dittmaier:1993bj,Dittmaier:1998nn,Weinzierl:2005dd}. 
At its core, it describes particles as (combinations of) two-component Weyl spinors which transform separately under Lorentz transformations.
At the level of the Lorentz algebra 
$so(3,1)_{\mathbb{C}} \cong \Lcolour{su(2)_{\mathbb{C},L}} \oplus \Rcolour{su(2)_{\mathbb{C},R}}$, 
the Weyl spinors transform under either the left-chiral 
$\Lcolour{su(2)_{\mathbb{C},L}}$ 
or the right-chiral $\Rcolour{su(2)_{\mathbb{C},R}}$.
For example, we can make this decomposition manifest by considering Dirac spinors in the chiral basis, 
written schematically (for some $p_1,p_2$) as
\begin{align}
u(p) &\sim v(p) \sim \begin{pmatrix}
\sqrSp{p_1} \\
\ranSp{p_2}
\end{pmatrix}
&
\ubar(p) &\sim \vbar(p) \sim \begin{pmatrix}
\sqlSp{p_1} \,\,\, , & \!\!\! \lanSp{p_2}
\end{pmatrix}
&
\gamma^5 &= \begin{pmatrix}
-1 & 0 \\
0 & 1
\end{pmatrix}
~,
\end{align}
where the 
\Lcolour{square} brackets are Weyl spinors transforming under $\Lcolour{su(2)_{\mathbb{C},L}}$,
the \Rcolour{angled} brackets are Weyl spinors transforming under $\Rcolour{su(2)_{\mathbb{C},R}}$,
and the eigenvalue of $\gamma^5$ gives the chirality.

Since helicity is the spin-quantum number of any massless particle 
\cite{Wigner:1939cj, Bargmann:1948ck, Weinberg:1995mt}, 
it is natural to calculate massless scattering amplitudes using states of definite helicity. 
For massless Weyl spinors, such states are also eigenstates of the chirality operator 
$\gamma^5$, meaning they transform under only one $su(2)$
\begin{alignat}{2}
    u^+(p) &= v^-(p) = \begin{pmatrix}
    0 \\
    \ranSp{p}
    \end{pmatrix} \qquad \qquad 
    u^-(p) &&= v^+(p) = \begin{pmatrix}
    \sqrSp{p} \\
    0
    \end{pmatrix}~, \nonumber \\
    \ubar^+(p) &= \vbar^-(p) = \begin{pmatrix}
    \sqlSp{p} \,\,\, , & \!\!\!  0
    \end{pmatrix} \qquad \,\,
    \ubar^-(p) &&= \vbar^+(p) = \begin{pmatrix}
    0 \,\,\, , & \!\!\!  \lanSp{p}
    \end{pmatrix}~.
    \end{alignat}  
Conversely, using $\tau^\mu = (1,\vec{\sigma})/\sqrt{2}$ and 
$\taubar^\mu = (1,-\vec{\sigma})/\sqrt{2}$,
vectors can be seen as containing both chiralities, 
with massless momenta given by
\begin{align}
\sqrt{2}p^\mu\tau_\mu &\equiv \slashed{p} = \sqrSp{p}\lanSp{p} ~, 
&
\sqrt{2}p^\mu\taubar_\mu &\equiv \bar{\slashed{p}} = \ranSp{p}\sqlSp{p}~,
\end{align} 
and outgoing gauge bosons given by \cite{Xu:1986xb,Gunion:1985vca}
\begin{align}
\eps_+^\mu(p,r) &= \frac{\lanSp{r}\taubar^\mu\sqrSp{p}}{\lan r p \ran}~,
&
\eps_-^\mu(p,r) &= \frac{\sqlSp{r}\tau^\mu\ranSp{p}}{\sql p r \sqr}~.
\end{align}
Here, the gauge boson has momentum $p$, while $r$
is an arbitrary reference momentum which corresponds to a particular gauge choice.

After using algebraic identities such as (see for example 
\cite{Dixon:1996wi,Elvang:2013cua})
\begin{align}
&\underbrace{\lanSp{i} \taubar^\mu \sqrSp{j} \sqlSp{k} \tau_\mu \ranSp{l} 
    = \lan il \ran \sql kj \sqr}_{
    \text{Fierz identity}}
& 
&\text{and}
&
&\underbrace{\lanSp{i} \taubar^\mu \sqrSp{j} =\sqlSp{j} \tau^\mu \ranSp{i}}_{
    \text{charge conjugation}}~,
\end{align}
a scattering amplitude is written in terms of Lorentz-invariant spinor inner products
\begin{align}
\lan ij \ran  &= -\lan ji \ran \equiv \lanSp{i} \ranSp{j} 
&
&\text{and}
&
    \sql ij \sqr &= - \sql ji \sqr \equiv \sqlSp{i} \sqrSp{j}~,
    &
    \lan ij \ran &\sim \sql ij \sqr \sim \sqrt{2p_i\cdot p_j}
\end{align}
which are simple, well known complex numbers. 

\section{Chirality flow}
In the last section we saw that with a few algebraic identities one can move from Feynman rules to complex numbers. 
In this section we describe a set of flow rules which eliminates the need for explicitly using these identities,
which are instead built into the flow rules \cite{Lifson:2020pai,Lifson:2020oll,Alnefjord:2020xqr}.

We begin with an ansatz for the spinor inner products. 
Since the left- and right-chiral states transform separately under Lorentz transformations,
we require two distinct line types. 
Inspired by square brackets having dotted indices and angled brackets having undotted indices,
we use dotted (more accurately dashed) lines to refer to square inner products, 
and solid lines for angled inner products
\begin{align}
                 \lanSp{i}^{\al}\ranSp{j}_{\al} &\equiv \lan i j \ran = -\lan ji \ran
                           = \raisebox{-0.2\height}{\includegraphics[scale=0.4]{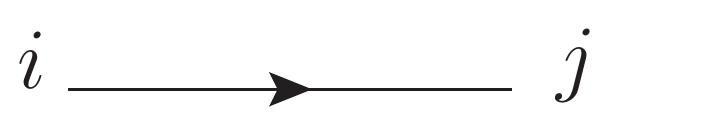}}, 
                 &
                 \sqlSp{i}_{\db}\sqrSp{j}^{\db} &\equiv \sql ij\sqr  = -\sql ji\sqr  = \raisebox{-0.2\height}{\includegraphics[scale=0.4]{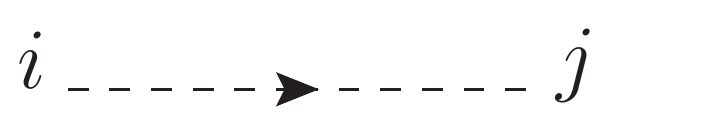}},        
        \end{align}
where the arrow direction matters since the inner products are antisymmetric.
Cutting the flow lines in two gives the flow definitions of the spinors
\begin{alignat}{2}
 \lanSp{i} &= \raisebox{-0.3\height}{\includegraphics[scale=0.5]{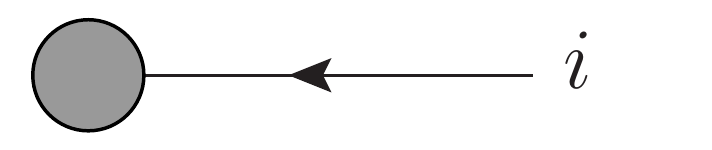}}~,
 \qquad \qquad 
   \sqlSp{i} &&= \raisebox{-0.3\height}{\includegraphics[scale=0.5]{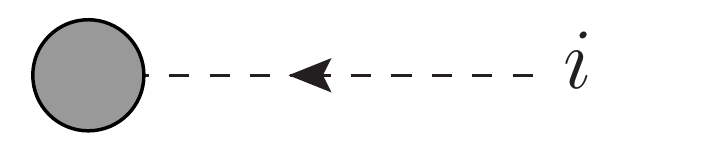}} ~,
 \nonumber \\
 \ranSp{j} &= \raisebox{-0.3\height}{\includegraphics[scale=0.5]{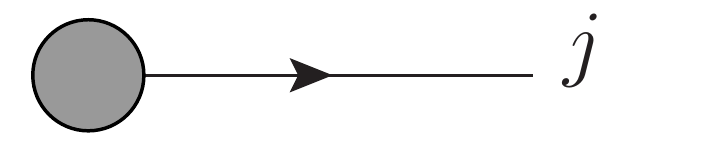}}~, 
   \qquad \qquad
   \sqrSp{j} &&= \raisebox{-0.3\height}{\includegraphics[scale=0.5]{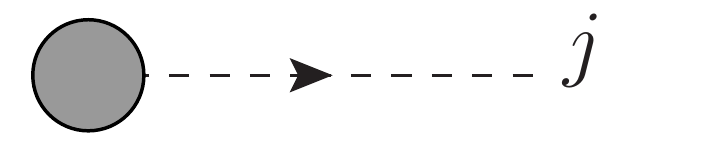}} ~.
\end{alignat}

In \cite{Lifson:2020pai}, we proved that we can always use the Fierz identity on the Pauli matrices, 
thus replacing a vector with a chirality-flow double line, i.e.\ 
a solid and dotted line with arrows opposing
\begin{equation}
\raisebox{-0.25\height}{\includegraphics[scale=0.45]{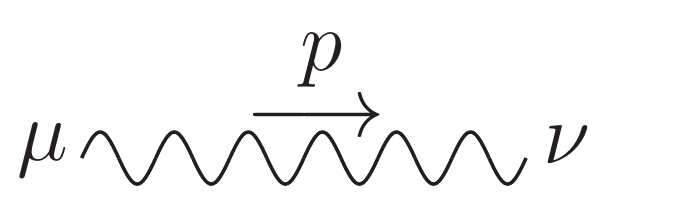}}  =  
\ \raisebox{-0.25\height}{\includegraphics[scale=0.45]{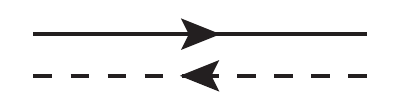}} \text{ or }
 \ \raisebox{-0.25\height}{\includegraphics[scale=0.45]{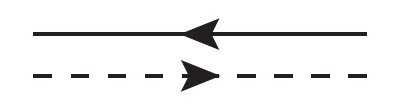}}~.
\label{eq:double line}
\end{equation}

Finally, we define the momentum dot for slashed momenta
(with $p=\sum_i p_i, \  p_i^2=0$)
\begin{align}
\sqrt{2}p^\mu\taubar_\mu &= \sum_i \ranSp{i}\sqlSp{i} = \raisebox{-0.25\height}{\includegraphics[scale=0.5]{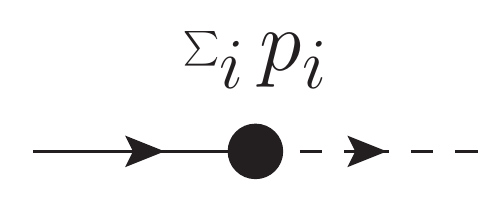}}~,
&
\sqrt{2}p^\mu\tau_\mu &= \sum_i \sqrSp{i}\lanSp{i} =  \raisebox{-0.25\height}{\includegraphics[scale=0.5]{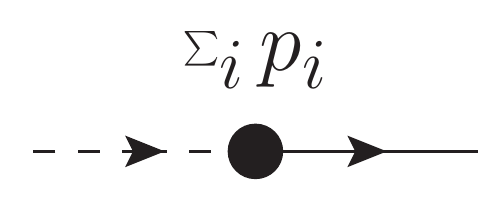}}~.
\end{align}

If the particles are massive, 
we describe them as combinations of massless spinors allowing to recycle the above results.
For instance, a massive momentum $p$ with $p^2=m^2\neq 0$ 
is decomposed as a sum of massless momenta $p^{\flat}$ and $q$
\begin{equation}
p^{\mu} = p^{\flat,\mu}+\alpha q^\mu~, \qquad (p^\flat)^2=q^2=0, \qquad p^2=m^2~,
\qquad \alpha = \frac{m^2}{2p^\flat\cdot q} = \frac{m^2}{2p\cdot q}~,
\end{equation}
while, for example, an incoming spinor with spin along the axis 
$s^\mu = (p^\mu - 2\alpha q^\mu)/m$ is given by
\begin{align}
u^+(p) &= 
\begin{pmatrix} -e^{-i\varphi}\sqrt{\alpha}\sqrSp{q} \\ 
        \phantom{-e^{-i\varphi}\sqrt{\al}}
        \ranSp{p^\flat} 
  \end{pmatrix}
= \begin{pmatrix}
-e^{\!-i\varphi}\!\sqrt{\alpha}
\raisebox{-5.5pt}{\includegraphics[scale=0.375]{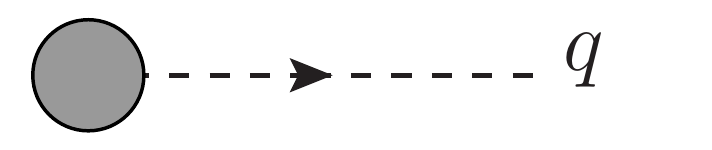}}
\hspace{-1.75ex}
\\
\phantom{-e^{\!-i\varphi}\!\sqrt{\alpha}}
\raisebox{-5.5pt}{\includegraphics[scale=0.375]{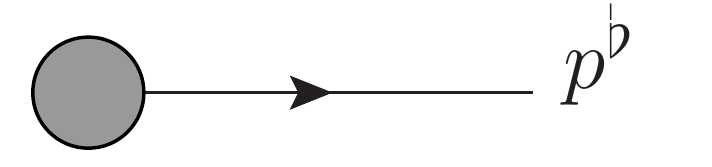}}
\hspace{-1.75ex}
\end{pmatrix}~,
&
e^{\!-i\varphi}\sqrt{\alpha}&=\frac{m}{\sql q p^\flat\sqr}~.
\label{eq:uPlus massive}
\end{align}
A full list of massive spinors and polarisation vectors,
together with the Standard Model flow rules is given in \cite{Alnefjord:2020xqr}.

	\section{Standard Model examples}
To calculate a Feynman diagram in massless QED, 
we simply draw the chirality-flow lines without the arrows, 
then connect them as given by the flow rules.
Next, we choose a single arrow direction and follow it through the diagram, 
remembering the requirement of opposing arrows for a double line, \eqref{eq:double line}. 
This process leads to an algebra-free journey from a Feynman diagram to inner products for even very complicated diagrams such as 
(Feynman in black, flow lines in colour, all momenta outgoing)

\vspace{-0.3cm}
\begin{minipage}{0.45\textwidth}
\hspace{-0.9cm}
\raisebox{-0.5\height}{\includegraphics[scale=0.6]{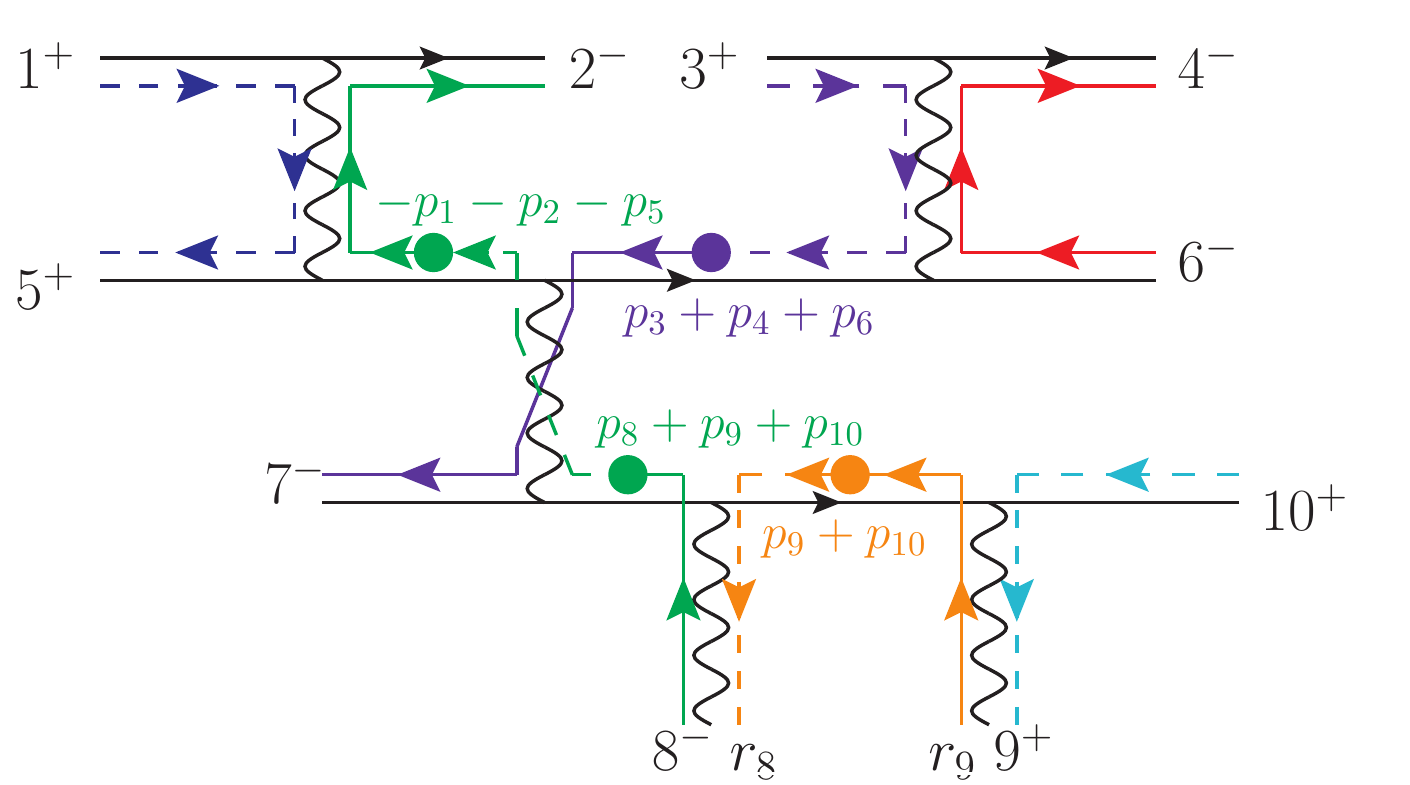}}
\end{minipage}   
\hspace{-0.4cm} 
\begin{minipage}{0.55\textwidth}
\begin{align*}
&=  
\underbrace{\frac{(-i)^3}{s_{1\,2}\;s_{3\,4}\;s_{7\,8\,9\,10}}}_{\text{photon propagators}} 
\underbrace{\frac{(i)^4}{s_{1\,2\,5}\;s_{3\,4\,6}\;s_{8\,9\,10}\;s_{9\,10}}}_{\text{fermion propagators}}
\underbrace{(\sqrt{2} e i)^8}_{\text{vertices}}
\\
& \times    
\underbrace{\frac{1}{[8 r_8 ] \langle r_9 9 \rangle}}_{\text{polarisation vectors}}
  {\textcolor{orange} {\Bigg(
    \langle r_9 9\rangle [9r_8]
    + \langle r_9 10\rangle [10r_8]
    \Bigg)}} 
\\
    &\times  {\textcolor{sky}{ [ 10\,\, 9]} }    {\textcolor{lilac}{\Bigg(
    \underbrace{[ 3 3]}_{0} \langle 3 7 \rangle
    + [3 4 ]\langle 4 7 \rangle
    + [ 3 6 ] \langle  6 7 \rangle
    \Bigg)}}
\end{align*}
\end{minipage}
\vspace{-0.3cm}
\begin{equation}
\times {\textcolor{blue}{[ 1 5 ]}}
  {\textcolor{red}{  \langle 6 4 \rangle} }
{\textcolor{jaxoGreen}{\Big(
      -\langle 8 9\rangle [9 1]\langle 1 2\rangle
      -\langle 8 9\rangle [9 5]\langle 5 2\rangle
      -\langle 8 \, 10\rangle [10\,\,1]\langle 1 2\rangle
      -\langle 8 \, 10\rangle [10\,\, 5]\langle 5 2\rangle
      \Big)}}~,
      \label{eq:QED example}
\end{equation}
where the flow line and inner product colours coincide and the black prefactors are trivially found.

When using massive fermions,
we have more components in our flow rules,
both from the external spinors (e.g.\ \eqref{eq:uPlus massive})
and from the mass term in the fermion propagator.
We then build the flow diagram from the flow rules as in \eqref{eq:QED example},
but have to take care of minus signs \cite{Alnefjord:2020xqr}.
For example, we find (ignoring trivial factors)
\begin{align}
&    \includegraphics[scale=0.40,valign=c]{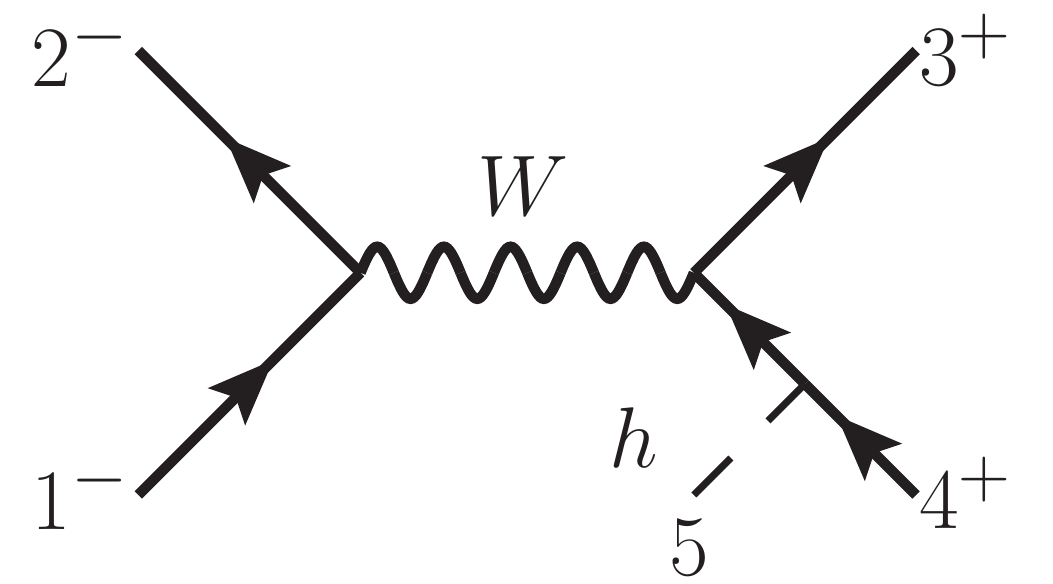} 
    \sim 
    \sqrt{\alpha_2\alpha_3} e^{i(\varphi_2+\varphi_3)} \nonumber \\
    &\times \left[\vphantom{ \includegraphics[scale=0.40,valign=c]{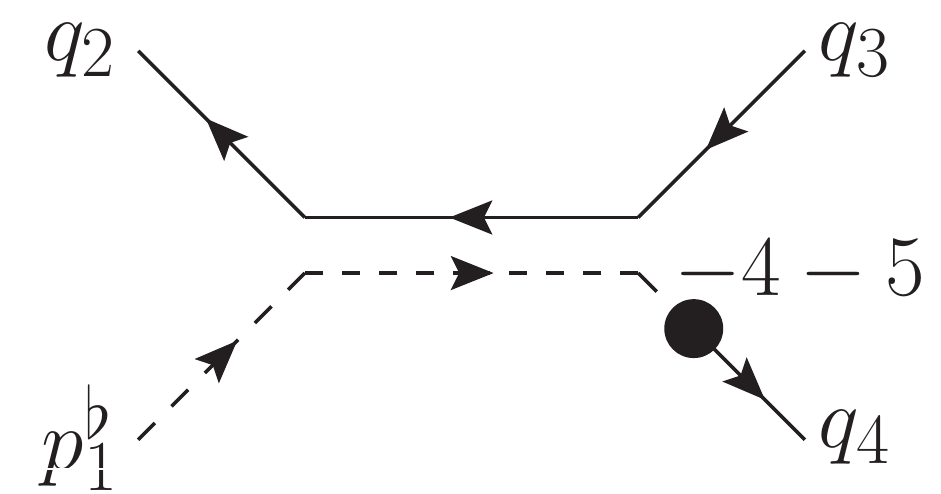}} \right.
    \sqrt{\alpha_4} e^{i\varphi_4}\includegraphics[scale=0.4,valign=c]{Jaxodraw/exampleMassiveChirFFFFHfullDot} \ - \  
    m_4 \includegraphics[scale=0.4,valign=c]{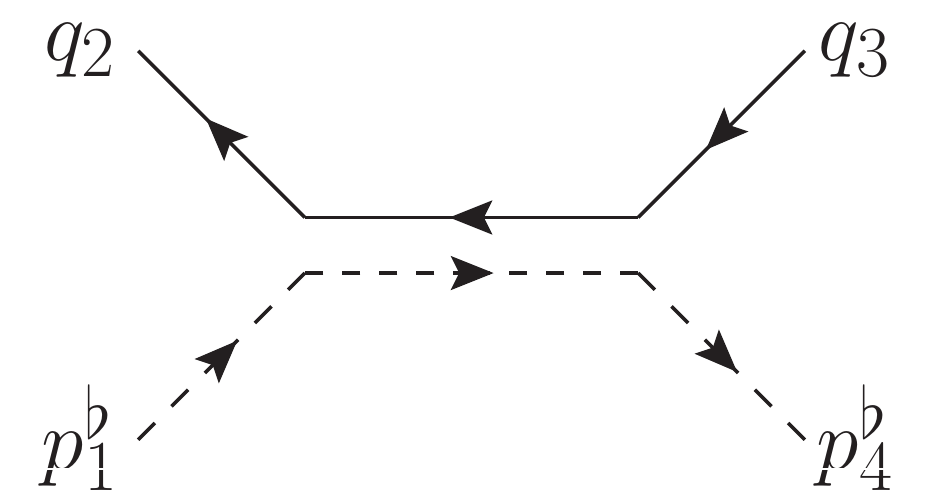}~
    \left. \vphantom{ \includegraphics[scale=0.4,valign=c]{Jaxodraw/exampleMassiveChirFFFFHfullDot}} \right]~,
    \end{align}
where the flow lines are the inner products we seek, 
the weak interaction simplifies by removing right-chiral couplings, 
and the Higgs has no flow since it is a Lorentz scalar.

\section{Conclusions and outlook}
In this presentation we reviewed the basics of the novel chirality-flow method, 
which allows to go from Feynman diagrams to complex numbers without intermediate algebraic manipulations.
We gave examples of massless and massive tree-level Feynman diagrams to illustrate the efficiency and transparency of our method,
which can be used to calculate any tree-level Standard Model process.
In future, we aim to extend this method to loops and recursive calculations.

\acknowledgments 
This work was supported by 
the Swedish Research Council (contract no.\ 2016-05996),
as well as the European Union's Horizon 2020 research and innovation programme
(grant agreement no.\ 668679).
This work has also received funding from the European Union's Horizon 2020 research and innovation programme as part of the Marie Sklodowska-Curie Innovative Training Network MCnetITN3 (grant agreement no.\ 722104).

\bibliographystyle{JHEP}
\bibliography{../paper_massive/chiralityflow_massive}

\providecommand{\href}[2]{#2}\begingroup\raggedright\begin{thebibliography}{10}

\bibitem{DeCausmaecker:1981jtq}
P.~De~Causmaecker, R.~Gastmans, W.~Troost and T.T.~Wu, \emph{{Multiple
  Bremsstrahlung in Gauge Theories at High-Energies. 1. General Formalism for
  Quantum Electrodynamics}},
  \href{https://doi.org/10.1016/0550-3213(82)90488-6}{\emph{Nucl. Phys.}
  {\bfseries B206} (1982) 53}.

\bibitem{Berends:1981rb}
F.A.~Berends, R.~Kleiss, P.~De~Causmaecker, R.~Gastmans and T.T.~Wu,
  \emph{{Single Bremsstrahlung Processes in Gauge Theories}},
  \href{https://doi.org/10.1016/0370-2693(81)90685-7}{\emph{Phys. Lett.}
  {\bfseries 103B} (1981) 124}.

\bibitem{Berends:1981uq}
F.A.~Berends, R.~Kleiss, P.~De~Causmaecker, R.~Gastmans, W.~Troost and T.T.~Wu,
  \emph{{Multiple Bremsstrahlung in Gauge Theories at High-Energies. 2. Single
  Bremsstrahlung}},
  \href{https://doi.org/10.1016/0550-3213(82)90489-8}{\emph{Nucl. Phys.}
  {\bfseries B206} (1982) 61}.

\bibitem{DeCausmaecker:1981wzb}
P.~De~Causmaecker, R.~Gastmans, W.~Troost and T.T.~Wu, \emph{{Helicity
  Amplitudes for Massless QED}},
  \href{https://doi.org/10.1016/0370-2693(81)91025-X}{\emph{Phys. Lett.}
  {\bfseries 105B} (1981) 215}.

\bibitem{Berends:1983ez}
{\scshape CALKUL} collaboration, \emph{{Multiple Bremsstrahlung in Gauge
  Theories at High-energies. 3. Finite Mass Effects in Collinear Photon
  Bremsstrahlung}},
  \href{https://doi.org/10.1016/0550-3213(84)90254-2}{\emph{Nucl. Phys.}
  {\bfseries B239} (1984) 382}.

\bibitem{Kleiss:1984dp}
R.~Kleiss, \emph{{The Cross-section for $e^+ e^- \to e^+ e^- e^+ e^-$}},
  \href{https://doi.org/10.1016/0550-3213(84)90197-4}{\emph{Nucl. Phys.}
  {\bfseries B241} (1984) 61}.

\bibitem{Berends:1984gf}
F.A.~Berends, P.H.~Daverveldt and R.~Kleiss, \emph{{Complete Lowest Order
  Calculations for Four Lepton Final States in electron-Positron Collisions}},
  \href{https://doi.org/10.1016/0550-3213(85)90541-3}{\emph{Nucl. Phys.}
  {\bfseries B253} (1985) 441}.

\bibitem{Gunion:1985bp}
J.F.~Gunion and Z.~Kunszt, \emph{{Four jet processes: gluon-gluon scattering to
  nonidentical quark - anti-quark pairs}},
  \href{https://doi.org/10.1016/0370-2693(85)90879-2}{\emph{Phys. Lett.}
  {\bfseries 159B} (1985) 167}.

\bibitem{Gunion:1985vca}
J.F.~Gunion and Z.~Kunszt, \emph{{Improved Analytic Techniques for Tree Graph
  Calculations and the G g q anti-q Lepton anti-Lepton Subprocess}},
  \href{https://doi.org/10.1016/0370-2693(85)90774-9}{\emph{Phys. Lett.}
  {\bfseries 161B} (1985) 333}.

\bibitem{Kleiss:1985yh}
R.~Kleiss and W.J.~Stirling, \emph{{Spinor Techniques for Calculating p anti-p
  {$\to$} W{$^{\pm}$} / Z$^0$ + Jets}},
  \href{https://doi.org/10.1016/0550-3213(85)90285-8}{\emph{Nucl. Phys.}
  {\bfseries B262} (1985) 235}.

\bibitem{Hagiwara:1985yu}
K.~Hagiwara and D.~Zeppenfeld, \emph{{Helicity Amplitudes for Heavy Lepton
  Production in e+ e- Annihilation}},
  \href{https://doi.org/10.1016/0550-3213(86)90615-2}{\emph{Nucl. Phys.}
  {\bfseries B274} (1986) 1}.

\bibitem{Kleiss:1986ct}
R.~Kleiss, \emph{{Hard Bremsstrahlung Amplitudes for $e^+ e^-$ Collisions With
  Polarized Beams at {LEP} / {SLC} Energies}},
  \href{https://doi.org/10.1007/BF01552550}{\emph{Z. Phys.} {\bfseries C33}
  (1987) 433}.

\bibitem{Kleiss:1986qc}
R.~Kleiss and W.J.~Stirling, \emph{{Cross-sections for the Production of an
  Arbitrary Number of Photons in Electron - Positron Annihilation}},
  \href{https://doi.org/10.1016/0370-2693(86)90454-5}{\emph{Phys. Lett.}
  {\bfseries B179} (1986) 159}.

\bibitem{Xu:1986xb}
Z.~Xu, D.-H.~Zhang and L.~Chang, \emph{{Helicity Amplitudes for Multiple
  Bremsstrahlung in Massless Nonabelian Gauge Theories}},
  \href{https://doi.org/10.1016/0550-3213(87)90479-2}{\emph{Nucl. Phys.}
  {\bfseries B291} (1987) 392}.

\bibitem{Gastmans:1987qz}
{\scshape CALKUL} collaboration, \emph{{New techniques and results in gauge
  theory calculations}},  in \emph{{Electroweak effects at high-energies.
  Proceedings, 1st Europhysics study conference, Erice, Italy, February 1-12,
  1983}}, pp.~599--609, 1987.

\bibitem{Schwinn:2005pi}
C.~Schwinn and S.~Weinzierl, \emph{{Scalar diagrammatic rules for Born
  amplitudes in QCD}},
  \href{https://doi.org/10.1088/1126-6708/2005/05/006}{\emph{JHEP} {\bfseries
  05} (2005) 006} [\href{https://arxiv.org/abs/hep-th/0503015}{{\ttfamily
  hep-th/0503015}}].

\bibitem{Farrar:1983wk}
G.R.~Farrar and F.~Neri, \emph{{How to Calculate 35640 O ($\alpha^5$) Feynman
  Diagrams in Less Than an Hour}},
  \href{https://doi.org/10.1016/0370-2693(85)90526-X,
  10.1016/0370-2693(83)91074-2}{\emph{Phys. Lett.} {\bfseries 130B} (1983)
  109}.

\bibitem{Berends:1987cv}
F.A.~Berends and W.~Giele, \emph{{The Six Gluon Process as an Example of
  Weyl-Van Der Waerden Spinor Calculus}},
  \href{https://doi.org/10.1016/0550-3213(87)90604-3}{\emph{Nucl. Phys.}
  {\bfseries B294} (1987) 700}.

\bibitem{Berends:1987me}
F.A.~Berends and W.T.~Giele, \emph{{Recursive Calculations for Processes with n
  Gluons}}, \href{https://doi.org/10.1016/0550-3213(88)90442-7}{\emph{Nucl.
  Phys.} {\bfseries B306} (1988) 759}.

\bibitem{Berends:1988yn}
F.A.~Berends, W.T.~Giele and H.~Kuijf, \emph{{Exact Expressions for Processes
  Involving a Vector Boson and Up to Five Partons}},
  \href{https://doi.org/10.1016/0550-3213(89)90242-3}{\emph{Nucl. Phys.}
  {\bfseries B321} (1989) 39}.

\bibitem{Berends:1988zn}
F.A.~Berends and W.T.~Giele, \emph{{Multiple Soft Gluon Radiation in Parton
  Processes}}, \href{https://doi.org/10.1016/0550-3213(89)90398-2}{\emph{Nucl.
  Phys.} {\bfseries B313} (1989) 595}.

\bibitem{Berends:1989hf}
F.A.~Berends, W.T.~Giele and H.~Kuijf, \emph{{Exact and Approximate Expressions
  for Multi - Gluon Scattering}},
  \href{https://doi.org/10.1016/0550-3213(90)90225-3}{\emph{Nucl. Phys.}
  {\bfseries B333} (1990) 120}.

\bibitem{Dittmaier:1993bj}
S.~Dittmaier, \emph{{Full O(alpha) radiative corrections to high-energy Compton
  scattering}}, \href{https://doi.org/10.1016/0550-3213(94)90139-2}{\emph{Nucl.
  Phys.} {\bfseries B423} (1994) 384}
  [\href{https://arxiv.org/abs/hep-ph/9311363}{{\ttfamily hep-ph/9311363}}].

\bibitem{Dittmaier:1998nn}
S.~Dittmaier, \emph{{Weyl-van der Waerden formalism for helicity amplitudes of
  massive particles}},
  \href{https://doi.org/10.1103/PhysRevD.59.016007}{\emph{Phys. Rev.}
  {\bfseries D59} (1998) 016007}
  [\href{https://arxiv.org/abs/hep-ph/9805445}{{\ttfamily hep-ph/9805445}}].

\bibitem{Weinzierl:2005dd}
S.~Weinzierl, \emph{{Automated computation of spin- and colour-correlated Born
  matrix elements}},
  \href{https://doi.org/10.1140/epjc/s2005-02467-6}{\emph{Eur. Phys. J.}
  {\bfseries C45} (2006) 745}
  [\href{https://arxiv.org/abs/hep-ph/0510157}{{\ttfamily hep-ph/0510157}}].

\bibitem{Wigner:1939cj}
E.P.~Wigner, \emph{{On Unitary Representations of the Inhomogeneous Lorentz
  Group}}, \href{https://doi.org/10.2307/1968551}{\emph{Annals Math.}
  {\bfseries 40} (1939) 149}.

\bibitem{Bargmann:1948ck}
V.~Bargmann and E.P.~Wigner, \emph{{Group Theoretical Discussion of
  Relativistic Wave Equations}},
  \href{https://doi.org/10.1073/pnas.34.5.211}{\emph{Proc. Nat. Acad. Sci.}
  {\bfseries 34} (1948) 211}.

\bibitem{Weinberg:1995mt}
S.~Weinberg, \emph{{The Quantum theory of fields. Vol. 1: Foundations}},
  Cambridge University Press (6, 2005).

\bibitem{Dixon:1996wi}
L.J.~Dixon, \emph{{Calculating scattering amplitudes efficiently}},  in
  \emph{{QCD and beyond. Proceedings, Theoretical Advanced Study Institute in
  Elementary Particle Physics, TASI-95, Boulder, USA, June 4-30, 1995}},
  pp.~539--584, 1996 [\href{https://arxiv.org/abs/hep-ph/9601359}{{\ttfamily
  hep-ph/9601359}}].

\bibitem{Elvang:2013cua}
H.~Elvang and Y.-t.~Huang, \emph{{Scattering Amplitudes}},
  \href{https://arxiv.org/abs/1308.1697}{{\ttfamily 1308.1697}}.

\bibitem{Lifson:2020pai}
A.~Lifson, C.~Reuschle and M.~Sjodahl, \emph{{The chirality-flow formalism}},
  \href{https://doi.org/10.1140/epjc/s10052-020-8260-8}{\emph{Eur. Phys. J. C}
  {\bfseries 80} (2020) 1006}
  [\href{https://arxiv.org/abs/2003.05877}{{\ttfamily 2003.05877}}].

\bibitem{Lifson:2020oll}
A.~Lifson, C.~Reuschle and M.~Sj\"odahl, \emph{{Introducing the Chirality-flow
  Formalism}}, \href{https://doi.org/10.5506/APhysPolB.51.1547}{\emph{Acta
  Phys. Polon. B} {\bfseries 51} (2020) 1547}.

\bibitem{Alnefjord:2020xqr}
J.~Alnefjord, A.~Lifson, C.~Reuschle and M.~Sjodahl, \emph{{The chirality-flow
  formalism for the standard model}},
  \href{https://doi.org/10.1140/epjc/s10052-021-09055-2}{\emph{Eur. Phys. J. C}
  {\bfseries 81} (2021) 371}
  [\href{https://arxiv.org/abs/2011.10075}{{\ttfamily 2011.10075}}].

\end{thebibliography}\endgroup

\end{document}